\documentclass[traditabstract,longauth]{aaemma} 
\usepackage{lineno}
\usepackage{graphicx,color}
\usepackage{natbib}
\usepackage{subfigure}
\bibpunct{(}{)}{;}{a}{}{,} 
\usepackage{graphicx,latexsym,amssymb}

\begin{document}

\newcommand\arcdeg{$^{\circ}$}
\newcommand\arcs{$^{\prime\prime}$}
\newcommand{\arcm}{$^\prime$}
\newcommand\snr{SNR\,G284.3--1.8}
\newcommand\psr{PSR\,J1016--5857}
\newcommand\src{HESS\,J1018--589}
\newcommand\srcfermi{1FGL\,J1018.6--5856}
\newcommand\xmmsrc{XMMU\,J101855.4--58564}
\newcommand\xmm{XMM-\emph{Newton}}

\title{Discovery of VHE emission towards the Carina arm region with the H.E.S.S. telescope array: HESS\,J1018--589}
\subtitle{}

\author{HESS Collaboration
\and A.~Abramowski \inst{1}
\and F.~Acero \inst{2}
\and F.~Aharonian \inst{3,4,5}
\and A.G.~Akhperjanian \inst{6,5}
\and G.~Anton \inst{7}
\and A.~Balzer \inst{7}
\and A.~Barnacka \inst{8,9}
\and Y.~Becherini \inst{10,11}
\and J.~Becker \inst{12}
\and K.~Bernl\"ohr \inst{3,13}
% ...
}

\institute{
Universit\"at Hamburg, Institut f\"ur Experimentalphysik, Luruper Chaussee 149, D 22761 Hamburg, Germany \and
Laboratoire Univers et Particules de Montpellier, Universit\'e Montpellier 2, CNRS/IN2P3,  CC 72, Place Eug\`ene Bataillon, F-34095 Montpellier Cedex 5, France \and
Max-Planck-Institut f\"ur Kernphysik, P.O. Box 103980, D 69029 Heidelberg, Germany \and
Dublin Institute for Advanced Studies, 31 Fitzwilliam Place, Dublin 2, Ireland \and
National Academy of Sciences of the Republic of Armenia, Yerevan  \and
Yerevan Physics Institute, 2 Alikhanian Brothers St., 375036 Yerevan, Armenia \and
Universit\"at Erlangen-N\"urnberg, Physikalisches Institut, Erwin-Rommel-Str. 1, D 91058 Erlangen, Germany \and
Nicolaus Copernicus Astronomical Center, ul. Bartycka 18, 00-716 Warsaw, Poland \and
CEA Saclay, DSM/IRFU, F-91191 Gif-Sur-Yvette Cedex, France \and
Astroparticule et Cosmologie (APC), CNRS, Universit\'{e} Paris 7 Denis Diderot, 10, rue Alice Domon et L\'{e}onie Duquet, F-75205 Paris Cedex 13, France \thanks{(UMR 7164: CNRS, Universit\'e Paris VII, CEA, Observatoire de Paris)} \and
Laboratoire Leprince-Ringuet, Ecole Polytechnique, CNRS/IN2P3, F-91128 Palaiseau, France \and
Institut f\"ur Theoretische Physik, Lehrstuhl IV: Weltraum und Astrophysik, Ruhr-Universit\"at Bochum, D 44780 Bochum, Germany \and
Institut f\"ur Physik, Humboldt-Universit\"at zu Berlin, Newtonstr. 15, D 12489 Berlin, Germany
% ...
}

\newpage

\offprints{\\emma@mpi-hd.mpg.de\\rterrier@apc.univ-paris7.fr}

\date{Received  ; accepted }
 
\abstract 
{The Carina arm region, containing the supernova remnant
  \object{SNR\,G284.3--1.8}, the high-energy (HE; E$>$100 MeV) binary
  \object{1FGL\,J1018.6--5856} and the energetic pulsar
  \object{PSR\,J1016--5857} and its nebula, has been
  observed with the H.E.S.S. telescope array. The observational
  coverage of the region in very-high-energy (VHE; E$>$0.1\,TeV) $\gamma$-rays benefits from
  deep exposure (40\,h) of the neighboring open cluster
  Westerlund~2. The observations have revealed a new extended region
  of VHE $\gamma$-ray emission. The
  new VHE source \object{HESS\,J1018--589} shows a bright, point-like
  emission region positionally coincident with \snr\ and \srcfermi\
  and a diffuse extension towards the direction of
  PSR\,J1016--5857. A soft ($\Gamma$=2.7$\pm$0.5$_{\rm stat}$) photon
  index, with a differential flux at 1\,TeV of
  N${\rm_0}$=(4.2$\pm$1.1)$\cdot$10$^{-13}$\,TeV$^{-1}$ cm$^{-2}$
  s$^{-1}$ is found for the point-like source, whereas the total
  emission region including the diffuse emission region is well fit by a
  power-law function with spectral index $\Gamma$=2.9$\pm$0.4$_{\rm
    stat}$ and differential flux at 1\,TeV of
  N${\rm_0}$=(6.8$\pm$1.6)$\cdot$10$^{-13}$\,TeV$^{-1}$ cm$^{-2}$
  s$^{-1}$. This H.E.S.S. detection motivated follow-up X-ray observations with the \xmm\
 satellite to investigate the origin of the VHE emission. The analysis
 of the \xmm\ data resulted in the discovery of a bright,
 non-thermal point-like source (XMMU J101855.4--58564) with a photon
 index of $\Gamma$=1.65$\pm$0.08 in the center of SNR\,G284.3--1.8, and
 a thermal, extended emission region coincident with its bright
 northern filament. The characteristics of this thermal emission are
 used to estimate the plasma density in the region as n$\approx$0.5
 cm$^{-3}$(2.9\,kpc/d)$^2$. The position of \xmmsrc\ is compatible with
 the position reported by the Fermi-LAT collaboration for the binary
 system \srcfermi\ and the variable Swift XRT source identified with
 it. The new X-ray data are used alongside archival multi-wavelength data
 to investigate the relationship between the VHE $\gamma$-ray emission
 from \src\ and the various potential counterparts in the Carina arm region.}

\authorrunning{H.E.S.S. Collaboration }
\titlerunning{H.E.S.S. detection of HESS\,J1018--589}

\maketitle

\section{Introduction}

\begin{figure}[t!]
 \centering
 \includegraphics[width=0.5\textwidth]{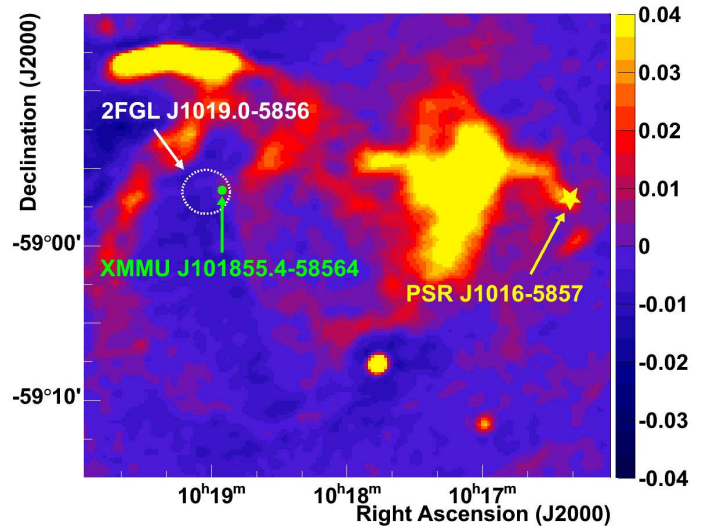}
 \caption{The MOST--Molonglo radio map (in units of Jy/beam) of \snr\ is displayed together
   with the position of the Fermi-LAT source 2FGL\,J1019.0--5856 (white dashed line showing the 95\% uncertainty in its position) and its
   associated X-ray source (green dot, XMMU J1018554--58564). To the west of the SNR
   the radio finger is visible extending toward the direction of \psr\
   (marked with a yellow filled star). The PSF size of the radio image is 45$^{\prime\prime}\times$52.5$^{\prime\prime}$.}
 \label{fig1}
\end{figure}

The H.E.S.S. (High Energy Stereoscopic System) collaboration has
carried out observations of the Carina arm as part of the Galactic
Plane Survey \citep{GPS1,GPS2,Gast}. The observed region includes three potential VHE $\gamma$-ray emitters, \snr\
(MSH 10--5$\it{3}$), the high spin-down luminosity pulsar \psr\
\citep{Camilo01} and the Fermi Large Area Telescope (Fermi-LAT)
$\gamma$-ray binary 1FGL\,J1018.6--5856 \citep{fermi1stcat}.
\snr\ has an incomplete radio shell (Fig. \ref{fig1}) with a non-thermal radio spectrum and a flux density of (5.4$\pm$0.8)\,Jy at 8.4\,GHz \citep{Milne89}. Evidence of
interaction with molecular clouds (MC; $^{12}$CO J=1--0) has been reported by \citet{Ruiz86} using observations with
the 1.2\,m Columbia Millimeter-Wave Telescope. The MC content
integrated over velocities in the local standard of rest from --14.95\,km\,s$^{-1}$ to --21.45\,km\,s$^{-1}$ traces
the radio shell shape of the SNR. 

The shell-like SNR shows a bright,
narrow filament, coincident with an H${\alpha}$ filament
\citep{Bergh73} to the East. Optical and CO observations imply that
\snr\ is most likely a 10$^{4}$\,yr old remnant of a type II supernova with a massive
stellar progenitor and that it is located at a distance of
d$_{SNR}$=2.9\,kpc (with an error of $\pm$20\%) \citep{Ruiz86}. The
western edge of the shell shows a so-called \emph{finger} emission
region (see Fig. \ref{fig1}) extended towards the direction of PSR\,J1016--5857. This Vela-like pulsar and its associated X-ray pulsar wind nebula (PWN), located $\approx$35\arcm\ away
from the geometrical center of the SNR, were discovered by \cite{Camilo01,Camilo04} in a search for counterparts of the unidentified source \object{3EG\,J1013--5915} \citep{EGRET}
with the Parkes telescope. The energetic pulsar, formerly
associated with the SNR\,G284.3--1.8, has a
rotation period of 107\,ms, a characteristic age $\tau_{\rm c}$=21\,kyr
and a spin-down luminosity of 2.6$\cdot$10$^{36}$\,erg\,s$^{-1}$. Its
distance can be estimated from the pulsar's dispersion measure to be
d$_{\rm PSR}$=9$^{+3}_{-2}$\,kpc \citep{TaylorCordes} or d$_{\rm PSR}$$\approx$8\,kpc \citep{CordesLazio}. Later observations with
{\it Chandra} \citep{Camilo04} resulted in the detection of the
associated X-ray PWN, apparently located at the tip of the finger of the SNR
radio emission. The PWN spectrum is well fit by a power-law model
with photon index $\Gamma_{\rm X}$=1.32$\pm$0.25, absorbing neutral
hydrogen column density N${\rm_H}$=(5.0$\pm$1.7)$\cdot$10$^{21}$\,cm$^{-2}$, and unabsorbed flux F$_{0.8-7 \rm keV}$=2.8$\cdot$10$^{-13}$\,erg\,cm$^{-2}$\,s$^{-1}$. The good match in both
width and position angle of the PWN and the SNR finger suggests a possible connection between the two. However, the association of
\psr\ with \snr\ is questioned by the large difference in their distance measurements and the large offset between the center of the shell and the pulsar position. 
\cite{Camilo04} argue that the uncertainties in the model for the free distribution of electrons in the Galaxy are large along spiral arm tangents, affecting the
estimation of the distance inferred from the dispersion
measure. A proper motion velocity of v$_{\rm
  PSR}$=500$\cdot$(d/3\,kpc)(21\,kyr/$\tau_{\rm c}$)\,km\,s$^{-1}$ has to be
invoked to connect the pulsar with the geometric center of the
SNR. Even if the proper motion velocity is within the range of known
pulsar velocities, this highly supersonic movement in the local
interstellar medium (ISM) would create a bow shock nebula, which
has not been observed.

The Fermi-LAT collaboration has reported the detection of a new source
1FGL\,J1018.6--5856 (or 2FGL\,J1019.0--5856), positionally coincident with \snr\ and associated
with the EGRET source 3EG\,J1013--5915 \citep{fermi1stcat,fermi2ndcat}. The source
reported in the LAT second-year source catalog shows an energy flux
of (3.53$\pm$0.11)$\cdot$10$^{-10}$\,erg\,cm$^{-2}$\,s$^{-1}$ in the
100\,MeV to 100\,GeV energy band, and a fit to a power-law spectrum
yields a photon index of $\Gamma{\rm_{HE}}$=2.59$\pm$0.04. The AGILE
collaboration also reported HE emission from this region and pulsed
emission is marginally detected (4.8$\sigma$) from the direction of PSR\,J1016--5857
\citep{AGILE}. The $\gamma$-ray emission from 1FGL\,J1018.6--5856,
however, shows a periodic modulation with a period of
16.58$\pm$0.02\,days, detected by Fermi-LAT
\citep{ATelFermi,fermiscience}. This periodicity is taken as evidence that
\srcfermi\ is a new $\gamma$-ray binary system, with an O6V((f))-type star (2MASS 10185560--5856459) proposed
as the stellar counterpart. The spectrum of the periodic source exhibits a break at $\approx$1\,GeV with best-fit values of $\Gamma{\rm_{HE}}$(0.1--1 GeV)=2.00$\pm$0.04 and $\Gamma{\rm_{HE}}$(1--10 GeV)=3.09$\pm$0.06 and an integral energy flux above 100\,MeV of (2.8$\pm$0.1)$\cdot$10$^{-10}$\,erg\,cm$^{-2}$\,s$^{-1}$. The binary nature of the HE source is
strengthened by the detection of a periodic X-ray (Swift XRT) and
radio (Australia Telescope Compact Array) compact source coincident
with the $\gamma$-ray source. A counterpart in hard X-rays has also
been detected using INTEGRAL data \citep{integral}.


\section{H.E.S.S. observations}
\label{sec2}
\begin{figure*}[t!]
 \centering
 \subfigure[]{
 \includegraphics[width=0.48\textwidth]{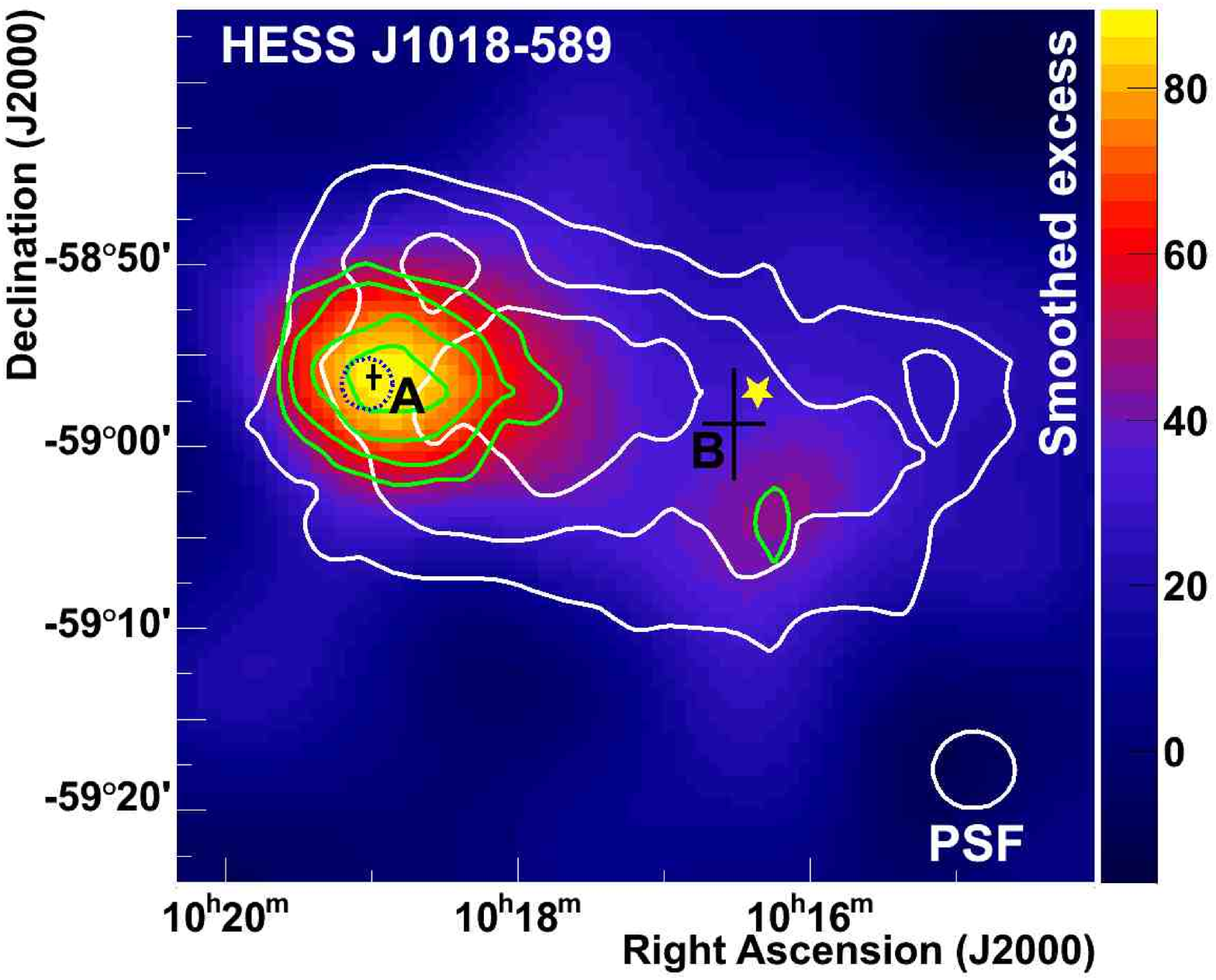}
\label{fig2a}
}
\subfigure[]{
  \includegraphics[width=0.48\textwidth]{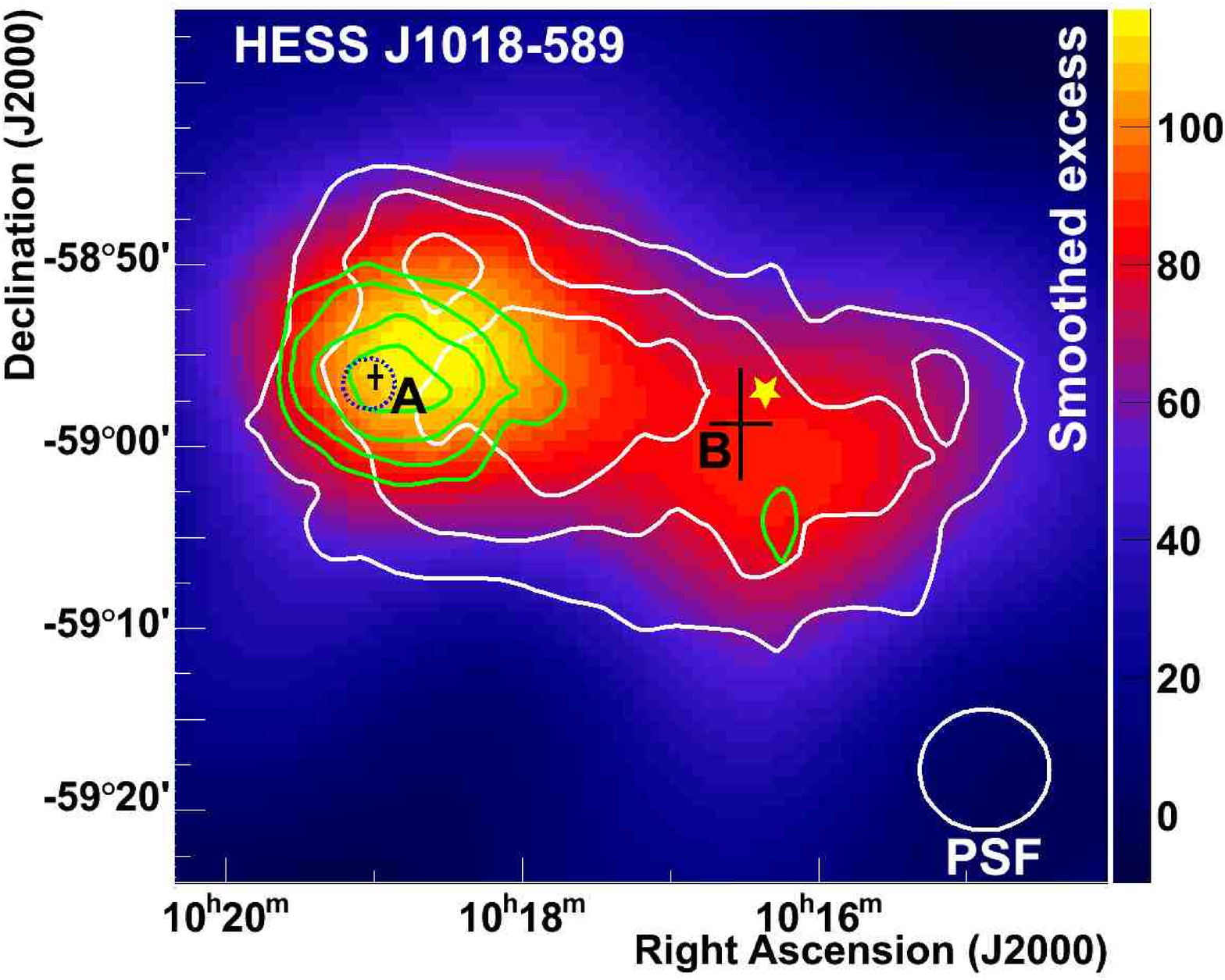}
\label{fig2b}
}
 \caption{H.E.S.S. excess image of the \src\ region smoothed with a
   Gaussian of width $\sigma$=0.07\arcdeg\ (on the left, a) and $\sigma$=0.11\arcdeg\ (on the right, b)). Significance contours, calculated using an oversampling radius of 0.1\arcdeg\ (in green) and 0.22\arcdeg\
   (in white) are shown, starting at 4$\sigma$ in steps of 1$\sigma$. The
   black crosses (A and B) mark the best-fit position and statistical errors
   at 1$\sigma$ level for the two emission regions (see text for details). The position of
   1FGL\,J1018.6--5856 is shown with a blue dashed ellipse (95\%
   confidence level) while the position of \psr\ is marked with a yellow
   star. The white circles illustrate the size of the (68\% containment radius) PSF smoothed
with the two different Gaussian widths used.}
  \label{fig2}
\end{figure*}

H.E.S.S. is an array of four VHE $\gamma$-ray imaging atmospheric
Cherenkov telescopes (IACTs) located in the Khomas Highland of Namibia
(23$^{\rm o}$16\arcm18\arcs\ S, 16$^{\rm o}$30\arcm00\arcs\ E). Each of these telescopes is equipped with a
tessellated spherical mirror of 107~m$^2$ area and a camera comprised
of 960 photomultiplier tubes, covering a large field-of-view (FoV) of
5$^{\rm o}$ diameter. The system works in a coincidence mode
(see e.g. \citealt{trigger}), requiring at least two of the four telescopes
to trigger the detection of an extended air shower (EAS). This
stereoscopic approach results in a high angular
resolution of $\approx$6\arcm\ per event, good energy resolution (15$\%$ on
average) and an efficient background rejection \citep{HESSCrab}. The mean (energy-dependent) point spread function (PSF) is estimated to be $\approx$0.1\arcdeg.
These characteristics
allow H.E.S.S. to reach a sensitivity of $\approx$2.0$\cdot$10$^{-13}$~ph~cm$^{-2}$s$^{-1}$ (equivalent to 1$\%$
of the Crab Nebula flux above 1\,TeV), or better if advanced techniques are used for image
analysis \citep{NGC253}, for a point-like source detected at a
significance of 5$\sigma$ in 25\,hours of observation at zenith.

The Carina arm region benefits from deep H.E.S.S. observations of the
Westerlund~2 region \citep{W2} and an acceptance-corrected effective exposure time
of 40\,h was obtained towards the direction of \snr\ and \psr\ in a multi-year observation campaign in January and March 2007, April and May 2008, and May to June 2009. The data set consists of scan-mode observations and dedicated observations in wobble-mode, in which the telescopes are pointed offset from the nominal source location to allow simultaneous background estimation (with an offset range in the total data set from 0.5\arcdeg\ to 2.6$^{\rm o}$). The observations were performed in a zenith angle range from 35\arcdeg\ to 50\arcdeg\ and the optical response of the system was estimated from the Cherenkov light of single muons as explained in \cite{HESSCrab}. 

The data have been analyzed using a multivariate analysis \citep{mva}
and cross-checked with the Hillas second moment \citep{Hillas} event reconstruction scheme
\citep{HESSCrab} and Model Analysis \citep{modelana}, including independent calibration of pixel
amplitudes and identification of problematic or dead pixels in the
IACTs cameras, leading to compatible results. To study the morphology
of the source with sufficient statistics, a cut of 80 photoelectrons
(p.e.) on the intensity of the EAS is used to reduce the data and
produce the images, light curve and spectrum of the source. The energy
threshold of the analysis presented here is E$_{\rm th}\approx$0.6\,TeV.

\begin{figure}[t!]
 \centering
 \includegraphics[width=0.5\textwidth]{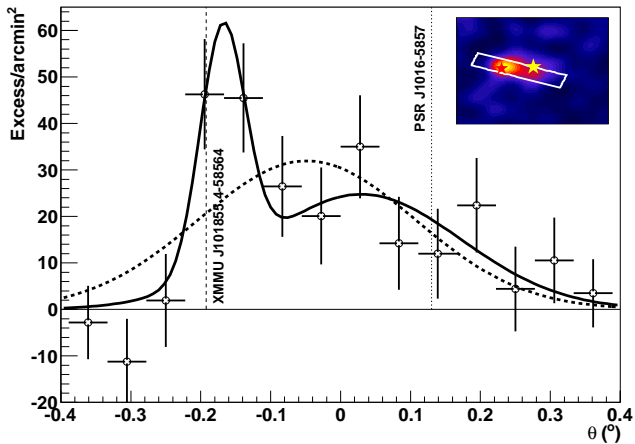}
 \caption{Profile of the VHE emission along the line between the peak
   of the point-like emission and the peak of the diffuse emission, as
   illustrated in the inset. Fits using a single and a double
   Gaussian function are shown in dashed and solid lines
   respectively. The positions of \xmmsrc\ and \psr\ are marked with
   dashed and dotted vertical lines and red and yellow stars in the inset, in which the significance image obtained using an oversampling radius of 0.1\arcdeg\ is shown.}
  \label{fig3}
\end{figure}

Figure \ref{fig2} a) shows an image of \src\ smoothed with a Gaussian of
width $\sigma$=0.07$^{\circ}$. The background in each pixel is
estimated with the ring background method \citep{HESSCrab}. This
image shows a point-like $\gamma$-ray excess centered on the position
of \srcfermi\ and SNR\,G284.3--1.8, and a diffuse emission region extending
in the direction of PSR\,J1016--5857. A peak pre-trials significance of
8.3$\sigma$ (which corresponds to over 6 sigma post-trials) is obtained using an oversampling radius of
0.10\arcdeg\ (green contours in Fig. \ref{fig2}). Fig. \ref{fig2} b)
shows the same image but smoothed using a Gaussian of width
$\sigma$=0.11\arcdeg\ to match the significance contours calculated with an
oversampling radius of 0.22\arcdeg\ (optimized for extended sources,
\citealt{GPS1}), which are overlaid in white. An extension of the VHE emission
towards the West is observed when using the larger radius. The white
circles in Fig. \ref{fig2} illustrate the size of the PSF smoothed
with the two different Gaussian widths used.

 To investigate the morphology of \src\ different models were fit to the uncorrelated excess map using the Sherpa fitting package \citep{Sherpa}. The
maximum-likelihood-ratio (MLR) test, performed using a Gaussian-shape
function convolved with the H.E.S.S. PSF, finds a point-like source at 
$\alpha$=10$^{\rm h}$18$^{\rm m}$59.3$^{\rm s}\pm$2.4$^{\rm s}_{\rm
  stat}$ and $\delta$=--58$^{\rm o}$56\arcm10\arcs$\pm$36\arcs$_{\rm
  stat}$  (J2000), marked in black in Fig. \ref{fig2} (position A). The systematic
error is estimated to be 20\arcs\ per axis
\citep{AngularRes}. Including a second source in the MLR-test results
in an improvement of the test statistics by 29.8 (or $\approx$4.4$\sigma$
for a four degree of freedom $\chi^2$-test). The best-fit
position of the second source (position B) is located at $\alpha$=10$^{\rm h}$16$^{\rm m}$31$^{\rm
  s}\pm$12$^{\rm s}_{\rm stat}$ and $\delta$=--58$^{\rm
  o}$58\arcm48\arcs$\pm$3\arcm$_{\rm stat}$ (J2000) with an extent of $\sigma$=0.15$^{\rm o}$$\pm$0.03$^{\rm o}$$_{\rm stat}$.

The two-emission-regions hypothesis is strengthened when the projection
of a rectangular region on the uncorrelated excess map along the extension
of the image (with a width of twice the H.E.S.S. mean PSF) is
evaluated. The fit of a single Gaussian function to the projection yields a $\chi^2$/$\nu$ of 21.04/13 (null hypothesis probability P=0.072) whereas the fit to a double
Gaussian function leads to a better $\chi^2$/$\nu$ of
9.6/10 (P=0.476). Fig. \ref{fig3} shows the profile of the rectangular region (in white in the inset). The single and double Gaussian function fits are
shown in dashed and solid lines, respectively.

\begin{figure}[t!]
 \centering
 \includegraphics[width=0.5\textwidth]{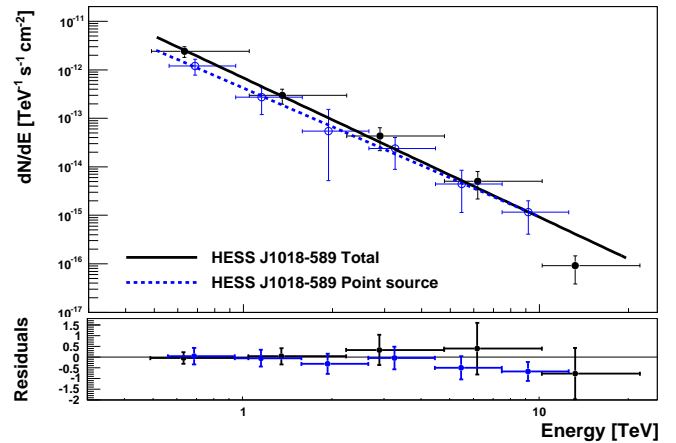}
 \caption{VHE photon spectrum of \src\ for a point-like source at position A (in blue dots and dashed blue line) and derived from a region of
   size 0.30\arcdeg\ comprising the point-like and diffuse emission (in
 black dots and solid black line.). The residuals to the fit are shown in the bottom panel.}
   \label{fig4}
\end{figure}

The energy spectrum (shown in Fig. \ref{fig4}) for the VHE excess was
computed by means of a forward-folding maximum likelihood fit
\citep{CATSpectrum}. To estimate the background the reflected background method was used, in which symmetric regions, not
contaminated by known sources, are used to extract the background
\citep{HESSCrab}. To derive the source spectrum two regions were selected, one around the point-like
source and a second region of 0.30\arcdeg\ radius accounting for the total
VHE $\gamma$-ray emission centered on $\alpha$=10$^{\rm h}$17$^{\rm m}$45.6$^{\rm
  s}$ and $\delta$=--59$^{\rm
  o}$00\arcm00\arcs (J2000). The photon spectra in both cases are well
represented with a simple power-law function
dN/dE=N${\rm_0}$(E/1\,TeV)$^{-\Gamma}$. The point-like source has a
photon index of $\Gamma$=2.7$\pm$0.5$_{\rm stat}\pm$0.2$_{\rm sys}$ statistically compatible with
the total (0.30$^{\rm o}$) VHE $\gamma$-ray emission region,
$\Gamma$=2.9$\pm$0.4$_{\rm stat}\pm$0.2$_{\rm sys}$. The normalization constants at 1\,TeV are N${\rm_0}$=(4.2$\pm$1.1)$\cdot$10$^{-13}$\,TeV$^{-1}$cm$^{-2}$s$^{-1}$ and N${\rm_0}$=(6.8$\pm$1.6)$\cdot$10$^{-13}$\,TeV$^{-1}$cm$^{-2}$s$^{-1}$ for the point-like source and the total emission region, respectively. 
The systematic error on the normalization constant N$_{\rm 0}$ is estimated from simulated data to be 20\% \citep{HESSCrab}.

\begin{figure}[t!]
 \centering
 \includegraphics[width=0.5\textwidth]{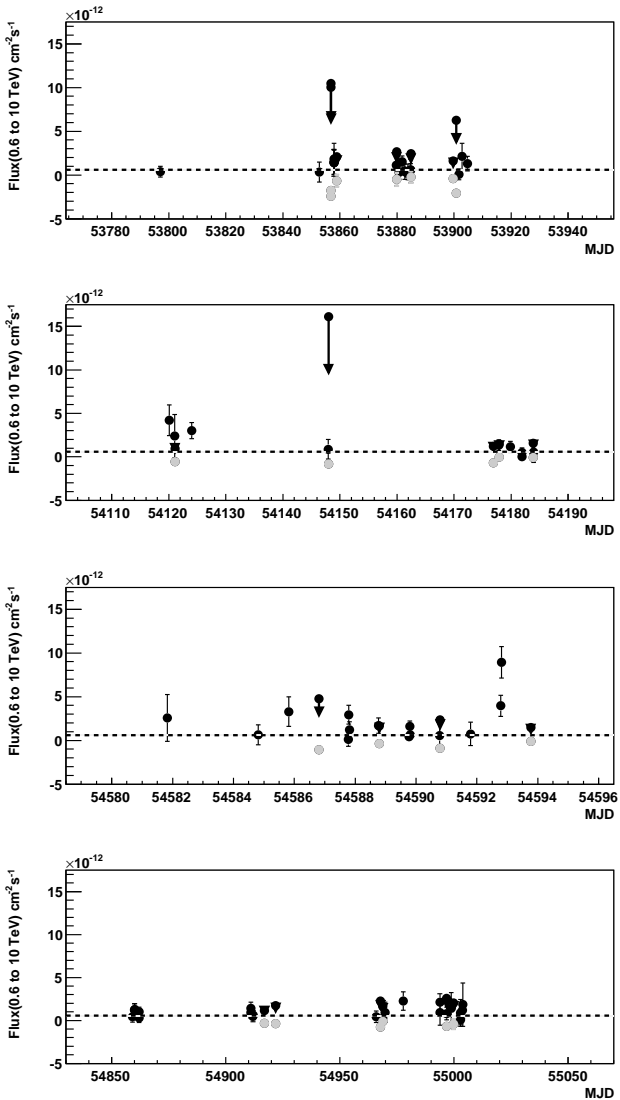}
 \caption{Light curve of the integral flux between 0.6\,TeV and 10\,TeV in a 0.1\arcdeg\ region centered on position A binned such that each point represents the data in one H.E.S.S observing run (typically 28 min). An upper limit for 99\% confidence level is shown with an arrow when the fluctuation in the integral flux is negative (grey points). The dashed horizontal line shows the mean integral flux.}
  \label{fig5}
\end{figure}

Finally, Fig. \ref{fig5} shows the light curve of the point-like emission region,
integrated between 0.6\,TeV and 10\,TeV. The Lomb-Scargle test \citep{LS}
was applied to the data to search for periodicity or variability
without positive results for this data set. Fitting a constant to the integrated flux yields a $\chi^2$/$\nu$ of 67.2/66, equivalent to a variable
integral flux on a run-by-run basis at the level of 1.0$\sigma$. The
lack of orbital coverage prevents any firm conclusion on
variability of the TeV emission at the Fermi-LAT reported period
(16.58\,days).

\section{X-ray observations with XMM-\emph{Newton}}
\label{sec3}

\xmm\ observations were acquired (ID: 0604700101, PI. E. de O\~na Wilhelmi) to investigate the
origin of the VHE emission region. The field was observed on the
22$^{\rm nd}$ August 2009,  with a total integration time of
20\,ksec. The observations were centered on  $\alpha$=10$^{\rm h}$18$^{\rm m}$55.60$^{\rm s}$ and $\delta$=--58\arcdeg55\arcm56.8\arcs (J2000) and acquired with the EPIC-PN
\citep{Struder2001} and EPIC-MOS \citep{Turner2001} cameras in
full-frame mode with a medium filter in a single pointing. This
position allowed a good coverage of the whole SNR structure in the
EPIC cameras.

The data were analyzed using the XMM Science Analysis System (SAS
v11.0.0 \footnote{http://xmm.esac.esa.int/sas/}) and calibration files
valid at September 2011. To exclude high background flares, which could
potentially affect the observations, light curves were
extracted above 10\,keV for the entire FoV of the EPIC cameras, but no
contamination was found. Therefore the full data set was used for the
image and spectral analysis. To create images, spectra, and light
curves, events with FLAG=0, and PATTERN=12 (MOS) and 4 (PN) were
selected. Hereafter clean event files in the 0.3 to 8\,keV energy band are used.

\begin{figure*}[t!]
 \centering
 \subfigure[]{
  \includegraphics[width=0.48\textwidth]{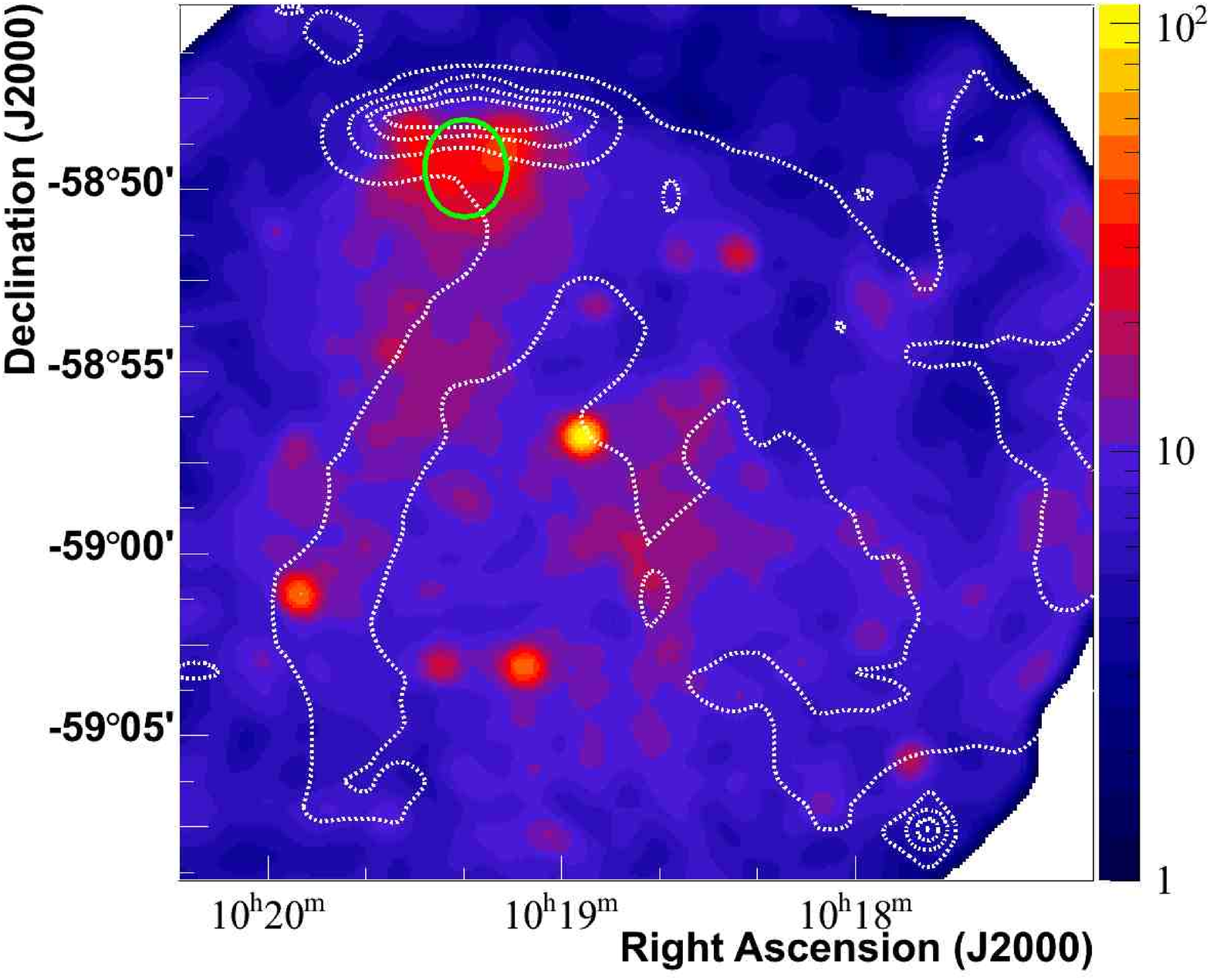}
\label{fig6a}
}
\subfigure[]{
 \includegraphics[width=0.48\textwidth]{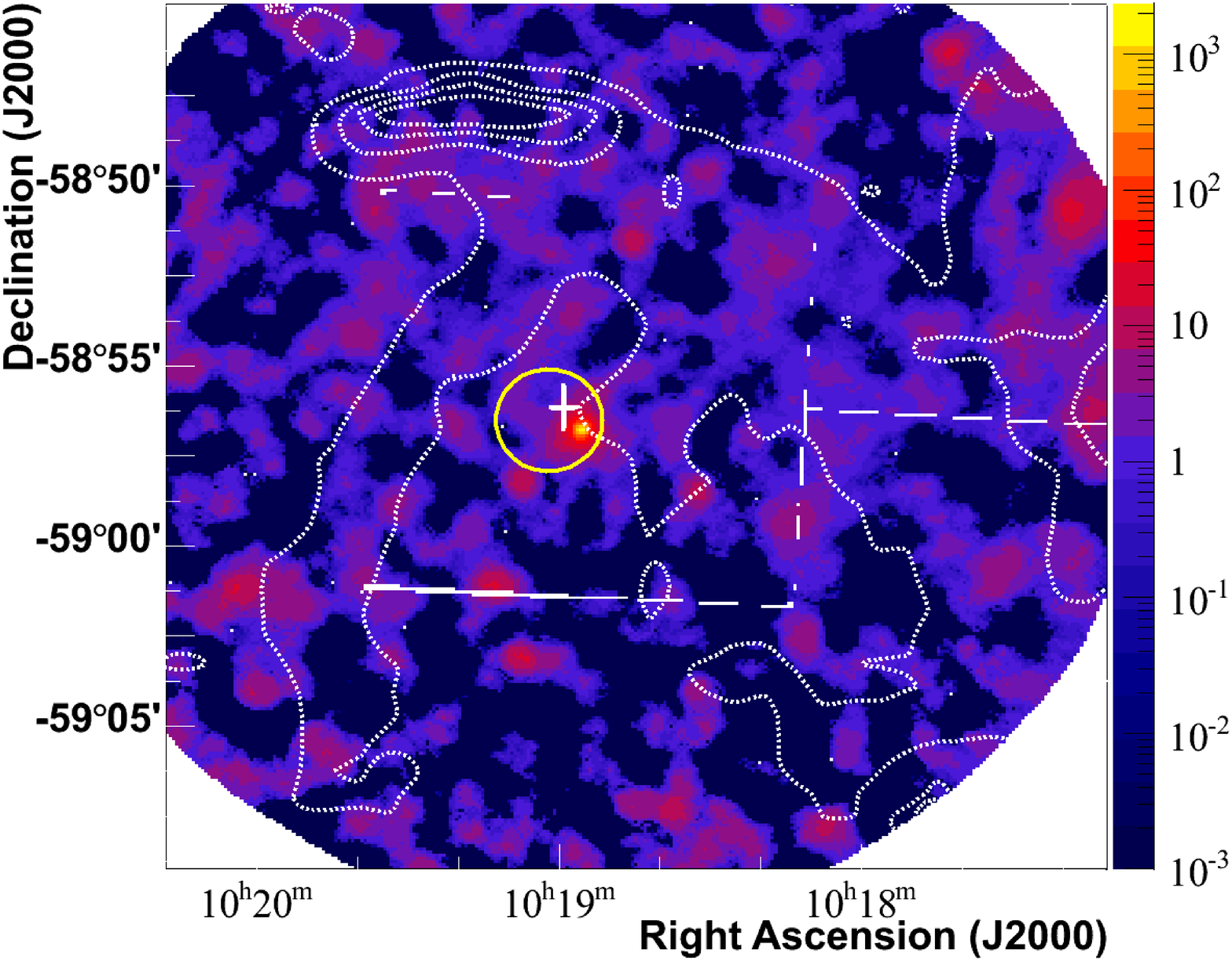}
\label{fig6b}
}
\subfigure[]{
 \includegraphics[width=0.48\textwidth]{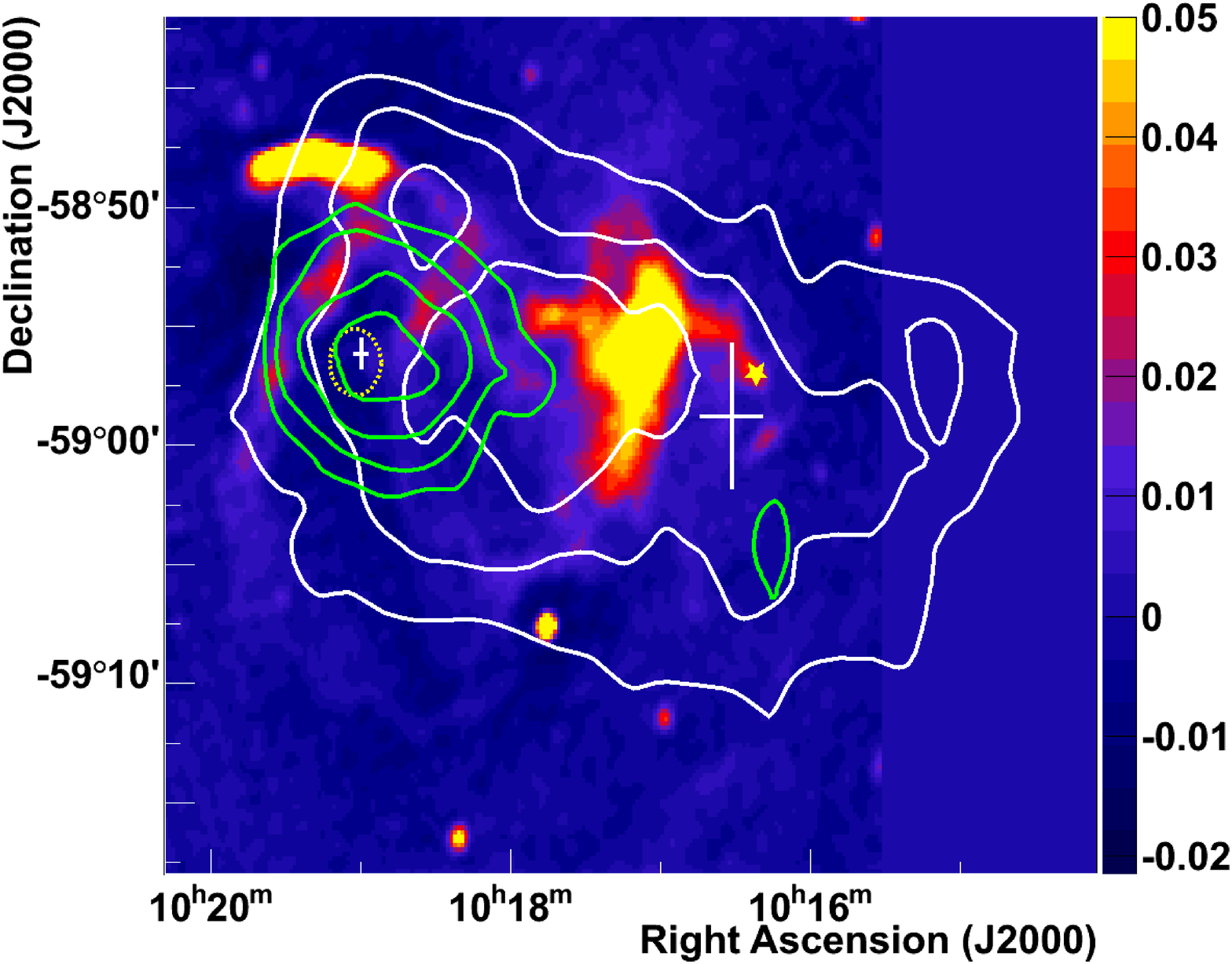}
\label{figc}
}
\subfigure[]{
  \includegraphics[width=0.48\textwidth,height=80mm]{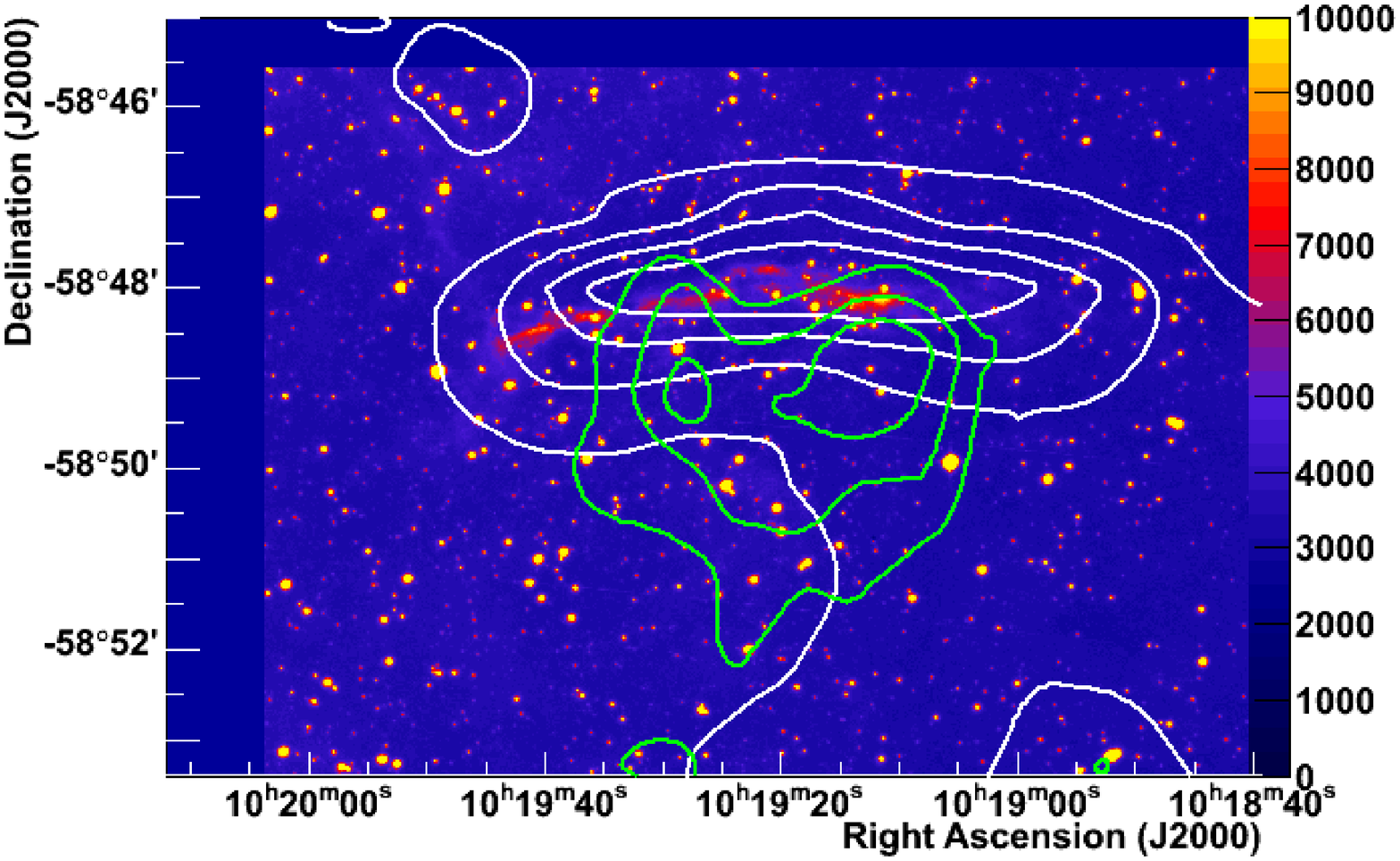}
\label{figd}
}
\caption{Multi-wavelength view of the \src\ region. On the top left a) 
   \xmm\ exposure-corrected counts map centered on \snr\ in the 0.3 to 2\,keV energy
   band in color scale combining the
   three EPIC cameras. The image (in excess per arcmin$^2$) is in
   logarithmic scale and it was adaptively smoothed to emphasize the extended thermal
   emission. The white-dashed contours show the radio image of \snr\ from
   MOST-Molonglo observations at 843\,MHz. The extraction region to
   evaluate the diffuse X-ray emission is shown in green. On the top right b) \xmm\ exposure-corrected counts map with the EPIC cameras at high energies in the 2 to
   8\,keV energy band. The white cross and yellow circle mark the best
   fit position of the H.E.S.S. point-like emission and the 95\%
   confidence level of the Fermi-LAT source 2FGL\,J1019.0--5856.  On the left bottom c) The significance contours for the two different oversampling radii for \src\ are shown in green (0.1$^{\circ}$) and white (0.22$^{\circ}$). The best-fit positions (white cross) of the VHE
   $\gamma$-ray, superimposed on the MOST-Molonglo radio map 
   are displayed, together with the position of \psr\ (yellow star)
   and 2FGL\,J1019.0--5856 (in dashed-yellow).  On the right bottom d) A zoom of
   the radio shell (in white) is shown together with the contours of
   the thermal emission in green and H$\alpha$ observations from the SuperCOSMOS H--alpha Survey (in intensity units).}
  \label{fig6}
\end{figure*}

Images combining the different EPIC instruments (see Fig. \ref{fig6} a)
and b)), vignetting-corrected and subtracted for particle induced and soft proton background, were
produced using the {\it ESAS} analysis package (integrated in SAS). In
the 0.3\,keV to 2\,keV energy range a bright source is detected (see Fig. \ref{fig6} a)). The radial profile of this central source was derived from the three EPIC cameras and fit with the
corresponding PSF, confirming its point-like nature within the
instrument angular resolution and observation sensitivity.  This
source is surrounded by diffuse emission extending up to the radio
shell. A strong enhancement of the diffuse emission is visible just
downstream of the radio and H${\alpha}$ filament (see Fig. \ref{fig6}
a) and d)). At higher energies (2 to 8\,keV, see Fig. \ref{fig6} b)), the
diffuse emission is strongly suppressed suggesting a thermal nature
of the emission pervading the SNR, the only significant feature being
the bright point-like central source.

\begin{figure}[t!]
 \centering
 \includegraphics[angle=-90,width=0.5\textwidth]{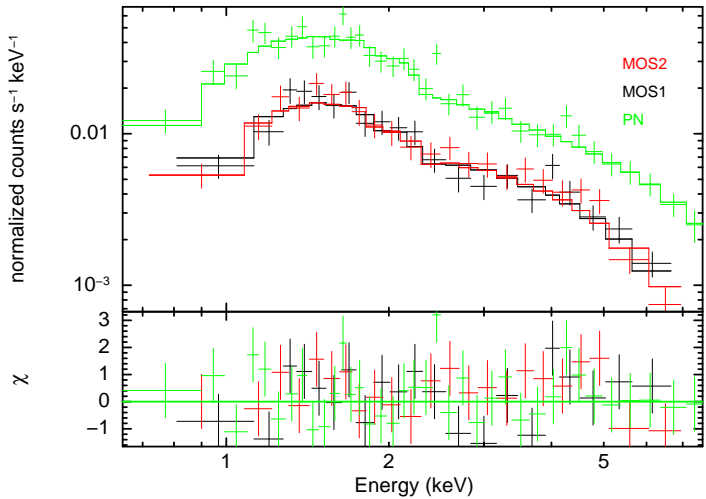}
 \caption{XMM-\emph{Newton}/PN/MOS spectrum of the point-like source
   XMMU J101855.4--58564, fit with an absorbed power-law model in the
   0.5 to 7.5\,keV energy range. The bottom panel shows the residuals to
   the fitted spectra.}
  \label{fig7}
\end{figure}

This source, dubbed \object{XMMU J101855.4--58564}, is located at $\alpha$=10$^{\rm
  h}$18$^{\rm m}$55.40$^{\rm s}$ and
$\delta$=--58\arcdeg56\arcm45.6\arcsec\ (J2000) with statistical error of
$\pm$0.25\arcsec\ in each coordinate (derived using the SAS task
\emph{edetect}). The position is compatible with the one derived by
\citet{Pavlov} The photon spectrum of the point-like source was derived integrating over a 20\arcsec\ circle around the fit position and the background was estimated from a circle of 40\arcsec\ located in the vicinity. The spectrum is well fit ($\chi^2_{\rm red}$=0.97, $\nu$=159) by an
absorbed power-law function in the 0.5 to 7.5\,keV energy range,
with a photon index of $\Gamma_{\rm x}$=1.66$\pm$0.11$_{\rm stat}$ and an integrated
flux of F$_{2-10 \rm\,keV}$=(6.5$\pm$0.7$_{\rm stat}$)$\cdot$10$^{-13}$\,erg\,cm$^{\rm
  -2}$\,s$^{\rm -1}$. The absorption column density N$_{\rm H}$ is (6.6$\pm$0.8$_{\rm stat}$)$\cdot$10$^{21}$ cm$^{\rm -2}$, supporting a Galactic
origin of the source. Other models such as a black-body model give fits that
are statistically inadequate. Fig. \ref{fig7} shows the measured
spectrum for MOS1, MOS2 and PN (in black, red and green
respectively).

Archival 2MASS (Two Micron All Sky Survey) data of the region show a bright star (with
magnitudes J=10.44$\pm$0.02, H=10.14$\pm$0.02 and K=10.02$\pm$0.02)
dubbed 2MASS 10185560--5856459, located at $\alpha$=10$^{\rm
  h}$18$^{\rm m}$55.6$^{\rm s}$ and
$\delta$=--58\arcdeg56\arcm46\arcsec\ (J2000), 1.3\arcsec\ away from
XMMU\,J101855.4--58564, the likely counterpart in the binary
system. This source also appears in the USNO\footnote{http://www.usno.navi/mil/USNO/} catalog with magnitudes
B=12.76 and R=11.16. A distance of d$_{*}$=5.4$^{+4.6}_{-2.1}$\,kpc to the
2MASS star has been estimated through photometry by \cite{2MASS}.
The position of \xmmsrc\ is also in agreement with the variable
compact object detected by Swift XRT at $\alpha$=10$^{\rm  h}$18$^{\rm
  m}$55.54$^{\rm s}$ and $\delta$=--58\arcdeg56\arcm45.9\arcsec\
(J2000) and the Fermi-LAT source \srcfermi\ \citep{ATelFermi}.

\begin{figure}[t!]
 \centering
 \includegraphics[angle=-90,width=0.5\textwidth]{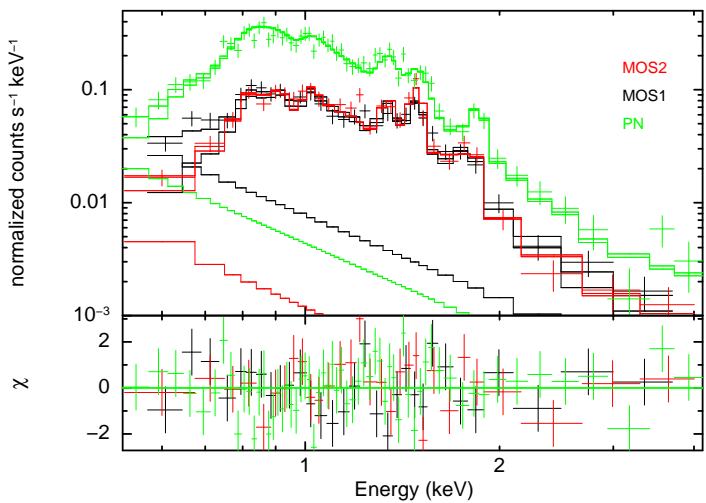}
 \caption{XMM-\emph{Newton}/PN/MOS spectrum of the extended emission region (in green in Fig.  \ref{fig6} a)) fit with an absorbed non-equilibrium ionization (PSHOCK) thermal model. The bottom panel shows the residuals to the fitted spectra.}
  \label{fig8}
\end{figure}

To the North-East of \xmmsrc\ a faint extended emission region located just
downstream of the radio and H${\alpha}$ filament of the remnant is
visible at low energy (E$<$2\,keV) (Fig. \ref{fig6} a) and d)). To extract the
X-ray spectrum, the background was modeled using the {\it ESAS}
software following the approach of \citet{arnaud01}. This background model is subsequently used to fit the signal
region. The final shell spectrum is well represented by an absorbed non-equilibrium ionization
(PSHOCK) thermal model
with a temperature of kT$\approx$0.5\,keV and a column density of 8$\cdot$10$^{21}$\,cm$^{\rm -2}$ (in Fig. \ref{fig8}). The normalization factor
A, defined as 
\begin{equation}
A (\rm cm^{-5}) = 10^{-14} \int n_e^2dV/4\pi d^2
\label{eq1}
\end{equation}
is 1.5$\cdot$10$^{-3}$\,cm$^{-5}$. From this value, assuming
a spherical volume (V) corresponding to the 2\arcmin\ source
extraction region (see Fig. \ref{fig6}a)) and a fully ionized gas
(n$_e$=1.2$\cdot$n), a plasma density can be derived to be n$\approx$0.5\,cm$^{-3}$(2.9\,kpc/d)$^2$.

\section{Discussion}
\label{sec4}

A new VHE $\gamma$-ray source, HESS\,J1018--589, has been discovered in the vicinity of
the shell-like remnant SNR\,G284.3--1.8. The extended TeV emission
region coincides with \snr\ and PSR\,J1016--5857, both viable candidates to explain
the observed VHE $\gamma$-ray emission. This emission region is also
positionally compatible with the new HE $\gamma$-ray binary \srcfermi\ reported
recently by the Fermi-LAT collaboration \citep{ATelFermi}.
Different possible scenarios in the context
of multi-wavelength observations, including new results derived from \xmm\ observations, are discussed in the following.

Although the spectral characteristics and light curve of the
H.E.S.S. source do not yet allow a firm identification of the origin
of the VHE $\gamma$-ray emission, the morphology of the source is
considered to clarify the situation. Two distinct emission regions are detected in the
H.E.S.S data, one point-like emission region (A) located in the center of
\snr\ with a centroid compatible with the  95\% confidence contour of
1FGL\,J1018.6--5856, and a diffuse emission region (B) extending towards
the direction of PSR\,J1016--5857, with its centroid compatible with the position of the pulsar.

The new HE $\gamma$-ray binary \srcfermi\ shares many characteristics with the VHE $\gamma$-ray binary LS\,5039 \citep{HESSLS5039}. The  \xmm\ observations presented here reveal a bright non-thermal point-like source, XMMU\,J101855.4--58564, in the center of the SNR and compatible with the position of the binary system and the H.E.S.S. point-like emission. The X-ray photon spectrum resembles that of pulsars, with a photon spectral index of 1.67 and a column
density of (6.6$\pm$0.8)$\cdot$10$^{21}$\,cm$^{\rm -2}$, compatible with that from the thermal
emission region coincident with the bright radio filament (7.9$\cdot$10$^{21}$\,cm$^{\rm -2}$). The position of \xmmsrc\
centered on the SNR and the similar distance (also compatible with the
distance to the associated 2MASS star) suggest a physical
association between the two objects (rather than an association between the SNR and PSR\,J1016--5857), namely that the compact object
within the binary system was the stellar progenitor for the type II SN
explosion (see e. g. \citealt{Bosch}). No extended
emission or putative PWN around the point-like source is observed in the present data at the level
of the \xmm\ observation sensitivity. The light curve of the X-ray
emission does not show any indication of variability or periodicity on short
time scales, for the time resolution of the MOS and PN cameras in
full-frame mode (2.6\,s and 73.4\,ms respectively).  \xmmsrc\ has been associated \citep{Pavlov}
with the Swift XRT source, which itself has been identified as a
counterpart to the HE binary system. The best-fit position of \src\
is compatible within less than 1$\sigma$ with both the position of
\xmmsrc\ and the variable HE source, whose position has been
determined accurately using timing analysis. The spectral type of the
possible companion star, O6V((f)) is similar to the one in the VHE
$\gamma$-ray binary LS\,5039. 

In a binary scenario composed of a massive star and a pulsar or a
black hole, modulated VHE $\gamma$-rays can be produced by different
mechanisms, namely inverse Compton emission or pions produced by
high-energy protons interacting with the stellar wind
\citep{bednarek,dubus,mitja,kirk,torres}. Similar to previously
detected VHE binary systems such as LS\,5039, LSI +61 303 \citep{lsi} and
PSR\,B1259--63 \citep{1259} periodic emission is found at
HE. Despite the similarities with other VHE binaries and in particular LS\,5039, and the good
positional agreement with the Fermi-LAT source, the association of the
H.E.S.S. source (A) with \srcfermi\ is still uncertain. No flux
variability has been observed yet and the Lomb-Scargle test does not
recover the Fermi-LAT reported 16.58 days modulation of the HE signal. Nevertheless it should be noted that the non-detection
might be due to possible contamination from the neighboring
diffuse emission, statistics of the data set, and and inadequate time
sampling of the orbit.

\src\ (A) is also coincident with the SNR\,G284.3--1.8.  SNRs are believed to be sites of particle acceleration up to at least a
few tens of TeV. Two types of VHE $\gamma$-ray emission associated with
SNRs have been discovered with IACTs, VHE $\gamma$-ray emission from
shell-like SNRs such as RX\,J1713.7--3947 \citep{j1713} or SN 1006 \citep{sn1006}, in which in general the VHE
morphology is in good agreement with the synchrotron X-ray emission;
and VHE $\gamma$-ray radiation which seems to originate through
proton-proton (p-p) interaction of cosmic rays (CR) accelerated in the
SNR interacting with local MCs in the vicinity, such as W28
\citep{w28}. \cite{Ruiz86} reported evidence of interaction of \snr\
with embedded MC, constraining the distance to the SNR to $\approx$2.9\,kpc. The analysis performed in the context of this work of public $^{\rm 12}$CO (J =1--0) data from the CfA
1.2~m Millimeter-Wave Telescope \citep{DameCO} yields an estimate of
the MC mass of $\approx$ 3$\cdot$10$^{3}$\,M$_{\odot}$. \cite{Ruiz86}
also reported on optical observations in the direction of \snr\ and
associated a bright optical filament (Fig. \ref{fig6} d)) coincident
with the brightest shell structure to the North-East, indicating
collisional excitation of the ISM, swept by the expanding SNR shock
wave. The observed MCs, if indeed physically associated with the SNR, could provide enough target material to
explain the VHE emission in a scenario in which the $\gamma$-ray are produced via p-p interaction. However, contrary to some other SNRs at VHE, the emission detected with
H.E.S.S. does not match the shell-type morphology within the present
statistics.

The diffuse emission detected with H.E.S.S. (B) extends towards the
direction of PSR\,J1016--5857. \psr\ was detected as a bright EGRET
source and pulsed emission at HE has been reported by the AGILE
\citep{AGILE} and the Fermi-LAT \citep{PulsarConference2010}
collaborations. With a spin-down luminosity of 2.6$\cdot$10$^{36}$\,erg\,s$^{-1}$, the radio,
HE and X-ray pulsar PSR\,J1016--5857 is energetic enough to power the
entire H.E.S.S. source, assuming a dispersion-measured estimated
distance of 9\,kpc. In this scenario, particles are accelerated in the
wind termination shock and produce VHE $\gamma$-ray emission by inverse Compton (IC) processes as they propagate away from the pulsar. As a result of
the interactions of relativistic leptons with the local magnetic field
and low-energy radiation, non-thermal radiation is produced up to
$\approx$100\,TeV (for a recent review see \citealt{pwn}).
Assuming a distance of 9\,kpc,  the total VHE luminosity in the 1 to 10\,TeV energy range is 9.7$\cdot$10$^{33}\cdot$(d/9\,kpc)$^2$\,erg\,s$^{-1}$, implying a maximum conversion from rotational energy into non-thermal emission with efficiency 0.4$\%$, with similar features to other well-established VHE PWNe, such as Vela~X \citep{VelaX} or HESS\,J1026--582 \citep{W2}. The associated X-ray nebula has been detected with Chandra in the 0.8 to 7\,keV energy range with a size of 3$^\prime$. 
The different size of the VHE and X-ray nebula,
$\approx$20$\cdot$(d/9\,kpc)\,pc and $\approx$8$\cdot$(d/9\,kpc)\,pc
respectively, can be easily accommodated in a relic nebula scenario,
and explained by the different energies (and hence cooling times) of
the electron population emitting X-rays and VHE $\gamma$-rays as seen,
e. g. in HESS\,J1825--137 \citep{HESSJ1825} for a low magnetic field of the order of a few $\mu$G.

\section{Conclusions}
\label{sec5}

A new VHE $\gamma$-ray source dubbed \src\ has been detected with the
H.E.S.S. telescope array with a significance of
8.3$\sigma$. The complex VHE morphology and faint VHE emission
prevent a unequivocal identification of the source given the presence of
several possible counterparts. The H.E.S.S. source seems to be composed of
two emission regions but the statistics are still too low to make firm
conclusions about the origin of those.

Several counterparts are discussed using energetics arguments as
to the possible origin of either part or all of the emission. In a SNR/MC scenario, \snr\ could partially
explain the VHE $\gamma$-ray emission via p-p interactions with the
associated MCs.  However the
fact that the VHE emission does not trace either the irregular shell or the
cloud morphology disfavors \snr\ as the only counterpart. 

The morphology and good positional agreement between the
H.E.S.S. best-fit position A and the new Fermi-LAT binary \srcfermi\
suggest a common origin. The analysis of the \xmm\ observations
revealed a non-thermal point-like source, XMMU\,J101855.4--58564, with photon spectral index of 1.67,
similar to the compact object found in LS 5039. Likewise, the spectral
class of the massive star companion listed in the 2MASS and USNO catalogs is similar to the one in LS\,5039. However, no variability has been found in the H.E.S.S. light curve. A dedicated observation campaign at VHE should help to clarify whether or not the two sources are indeed associated. 

The energetic pulsar PSR\,J1016--5857, also recently detected in Fermi-LAT and AGILE data, and
its X-ray nebula seem the most likely candidate to power the extended VHE $\gamma$-ray source,
given the high spin-down luminosity (2.6$\cdot$10$^{36}$\,erg\,s$^{\rm -1}$) and
X-ray nebula, which implies a population of high energy electrons able to up-scatter soft photon fields to VHE. The estimated age
of the pulsar (21\,kyr) would also explain the large size of the VHE nebula,
similar to other systems such as Vela X \citep{VelaX}.

Finally, \xmm\ observations also revealed thermal emission behind the
brightest synchrotron part of the radio shell of \snr, which might be associated with shock heated interstellar matter. The column density is statistically
compatible with the one derived from the direction of
XMMU\,J101855.4--58564. The similar column density and the position of
the pulsar candidate with respect to the center of the SNR could
indicate a common origin, where \xmmsrc\ is interpreted as the pulsar
left behind after the supernova explosion.

\begin{acknowledgements}
The support of the Namibian authorities and of the University of Namibia
in facilitating the construction and operation of H.E.S.S. is gratefully
acknowledged, as is the support by the German Ministry for Education and
Research (BMBF), the Max Planck Society, the French Ministry for Research,
the CNRS-IN2P3 and the Astroparticle Interdisciplinary Programme of the
CNRS, the U.K. Science and Technology Facilities Council (STFC),
the IPNP of the Charles University, the Polish Ministry of Science and 
Higher Education, the South African Department of
Science and Technology and National Research Foundation, and by the
University of Namibia. We appreciate the excellent work of the technical
support staff in Berlin, Durham, Hamburg, Heidelberg, Palaiseau, Paris,
Saclay, and in Namibia in the construction and operation of the
equipment.
This research has made use of the NASA/ IPAC Infrared Science Archive,
which is operated by the Jet Propulsion Laboratory, California
Institute of Technology, under contract with the National Aeronautics
and Space Administration.
\end{acknowledgements}


\end{document}